\documentclass[10pt,letterpaper]{article}

\usepackage{opex3}
\usepackage{color}
\usepackage{threeparttable}
\usepackage{epstopdf}
\usepackage{graphicx}
\usepackage{mathrsfs}
\usepackage{amssymb}
\usepackage{amsmath}
\usepackage{dsfont}
\usepackage{cite}

\begin{document}

\title{Population transfer driven by far-off-resonant fields}

\author{Z. C. Shi,$^{1,2}$ W. Wang,$^{2}$ and X. X. Yi$^{2,*}$}

\address{$^1$Department of Physics, Fuzhou University, Fuzhou 350002, China\\
$^2$Center for Quantum Sciences and School
of Physics, Northeast Normal University, Changchun 130024, China}

\email{$^*$yixx@nenu.edu.cn} 



\begin{abstract}
For a two-level system, it is believed that a far-off-resonant
driving can not help  coherent  population transfer between two
states. In this work, we propose a scheme to implement the coherent
transfer with far-off-resonant driving. The scheme works well with
both constant driving and Gaussian driving.
The total time to finish population transfer is also minimized by
optimizing the detuning and coupling constants. We find that the scheme
is sensitive to  spontaneous emission much more than dephasing. It might find
potential applications in X-ray quantum  optics and population
transfer in Rydberg atoms as well.
\end{abstract}




\section{Introduction}

Preparation and manipulation of a well-defined quantum state is  of
fundamental importance since it is extensively applied to many
branches of  physics ranging from laser physics \cite{scully97} to
quantum information processing \cite{nielsen10}.

In quantum information processing, using highly excited Rydberg
states allows one to switch on and off  strong interaction that are
necessary for engineering many-body quantum states, towards
implementing quantum simulation with large arrays of Rydberg atoms
\cite{saffman10}. However, the transition frequency from Rydberg
states to ground state falls  generally in the ultraviolet region,
which makes the transition difficult with a single laser. To
complete the transition, two lasers (two-photon processes) are
required \cite{schempp15,barredo15}. The drawback by this method is
that the transfer efficiency may not be high. This gives rise to a
question that does the transition from the ground state to Rydberg
states occur by lasers with frequency far-off-resonant with respect
to the transition frequency?

Long-lived excited nuclear states (also known as isomers) can store
large amounts of energy over longer periods of time.  Release on
demand of the energy stored in the metastable state called isomer
depletion together with nuclear batteries
\cite{walker99,ledingham03,aprahamian05,paffy07}, have received
great attention in the last one and a half decades. Depletion occurs
when the isomer is excited to a higher level, which is associated
with freely radiating states and therefore releases the energy of
the metastable state. Coherent population transfer between nuclear
states would therefore not only be a powerful tool for preparation
and detection in nuclear physics, but also especially useful for
control of energy stored in isomers.

In atomic physics, a successful and robust technique for atomic
coherent population transfer is the stimulated Raman adiabatic
passage (STIRAP) \cite{bergman98}. The transfer of such techniques
to nuclear systems, although encouraged by progress of laser
technology, has not been accomplished due to the lack of
$\gamma$-ray lasers.

To bridge the gap between X-ray laser frequency and nuclear
transition energies, a key scheme  is to combine moderately
accelerated target nuclei and novel X-ray lasers \cite{burvenich06}.
Using this proposal, the interaction of X-rays from the European
X-ray Free Electron Laser (XFEL) \cite{altarelli06} with nuclear
two-level systems was studied theoretically
\cite{burvenich06,palffy08}. However, the coherent control of
isomers have
 never been addressed, partially because of the poor
coherence properties of the XFEL. This raises  again the question
that whether the population transfer can be performed with high
fidelity by off-resonant drivings. In this work, we present a
proposal for  population transfer between two  states using lasers
off-resonant with the transition energies of the two states.
Note that there also exist several work \cite{baranov14,ber15,kempf15}
investigating population transfer by using only non-resonant driving
fields recently, which can be explained by the superoscillation phenomenon.
However, the major difference is that the driving fields is periodic in our work. In
particular, we consider two types of pulse, i.e., square-well
pulse and Gaussian pulse, as the driving fields.

\section{Physical mechanism for population transfer}

Consider a two-level system described by a Hamiltonian with
general form ($\hbar=1$)
\begin{eqnarray} \label{1}
H=\vec{d}\cdot\vec{\sigma}+\varepsilon\cdot\mathds{1},
\end{eqnarray}
where $\vec{d}=(d_x,d_y,d_z)$, and
$\vec{\sigma}=(\sigma_x,\sigma_y,\sigma_z)$  represent Pauli
matrices. $\mathds{1}$ is the $2\times2$ identity matrix. This
simple model can be employed to describe various physical  systems
ranging from natural microscopic system (e.g., atomic or spin
system) to artificial mesoscopic system (e.g., superconducting
circuits or semiconductor quantum dots). Since the
identity matrix term adds an overall energy level shift
to system, it only affects the global phase in  dynamics evolution and can
be ignored safely in later discussion.  The evolution operator of this system then can
be obtained after some straightforward algebras,
\begin{eqnarray} \label{2a}
U(t,0)=e^{-iHt}=\left(
         \begin{array}{cc}
          P(t)-iQ(t)  & -R(t)e^{-i(\theta-\frac{\pi}{2})} \\
           R(t)e^{i(\theta-\frac{\pi}{2})} & P(t)+iQ(t) \\
         \end{array}
       \right),
\end{eqnarray}
where $P(t)=\cos(|\vec{d}|t)$,
$Q(t)=\frac{d_z}{|\vec{d}|}\sin(|\vec{d}|t)$,
$R(t)=\frac{\sqrt{d_x^2+d_y^2}}{|\vec{d}|}\sin(|\vec{d}|t)$,
$|\vec{d}|=\sqrt{d_x^2+d_y^2+d_z^2}$, and
$\tan\theta=\frac{d_y}{d_x}$. When $d_z$ is sufficiently large
(i.e., $d_z\gg\sqrt{d_x^2+d_y^2}$), $R(t)\simeq0$. As a result, the
transition between two levels will be sharply suppressed.

In this work we focus on the system dynamics driven by periodic
square-well driving field. That is, the system is governed by the Hamiltonian
$H_1=\vec{d_1}\cdot\vec{\sigma}$ in the time interval $[0,t_1],$
while the Hamiltonian is $H_2=\vec{d_2}\cdot\vec{\sigma}$ in the
time interval $(t_1,T]$. $T$ is the period of the  square-well
driving field,  $t_2=T-t_1$. The evolution operator within  one
period of time can be written as,
\begin{eqnarray}
U(T,0)=e^{-iH_2t_2}e^{-iH_1t_1}=\left(
         \begin{array}{cc}
           \cos\phi & -\sin\phi e^{-i\theta} \\
           \sin\phi e^{i\theta} & \cos\phi \\
         \end{array}
       \right),
\end{eqnarray}
where
$\tan\phi=\frac{|\vec{v}\times\vec{d_1}\times\vec{d_2}|}{\vec{d_1}\cdot\vec{d_2}}$,
$\vec{v}=(0,0,1)$,
$\tan\theta=\frac{d_{1y}}{d_{1x}}=\frac{d_{2y}}{d_{2x}}$,  and
$|\vec{d_1}|t_1=|\vec{d_2}|t_2=\frac{\pi}{2}+2m\pi$ ($m$ is an arbitrary integer). For a periodic
system, one can calculate the time-independent effective Hamiltonian
from the evolution operator  via definition $U(T,0)\equiv
e^{-iH_{eff}T}$ \cite{eastham73}, and the effective Hamiltonian
reads
\begin{eqnarray}  \label{4a}
H_{eff}=\frac{1}{T}\left(
\begin{array}{cc}
0 & \phi e^{-i(\theta+\frac{\pi}{2})} \\
\phi e^{i(\theta+\frac{\pi}{2})} & 0 \\
\end{array}
\right).
\end{eqnarray}
It is interesting to find that in the periodic driving field,  the
transition between two levels is  determined by $\phi$ instead of
the condition $d_z\gg\sqrt{d_x^2+d_y^2}$ (it is called large
detuning condition in atomic system).
In other words, the periodic driving field can be viewed as the
constant driving field with the effective coupling strength
$\Lambda=\frac{\phi}{T}e^{-i(\theta+\frac{\pi}{2})}$ at
the exact resonance condition. Although the effective coupling strength may be very
small, it can still realize population transfer with the
increasing of evolution time.

Note that, at an arbitrary time $t=t'+nT$ ($n$ is the number of evolution periods),
the evolution operator can be written as
\begin{eqnarray}
U(t,0)=\left \{
\begin{array}{ll}
    e^{-iH_1t'}U(nT,0), & t'\in[0,t_1], \\
    e^{-iH_2(t'-t_1)}e^{-iH_1t_1}U(nT,0),& t'\in(t_1,T], \\
\end{array}
\right.
\end{eqnarray}
where the analytical expressions of $e^{-iH_1t'}$ and $e^{-iH_2t'}$ can be obtained from  Eq.(\ref{2a}),
and $U(nT,0)=\vec{d'}\cdot\vec{\sigma}+\cos n\phi\cdot\mathds{1}$,
$\vec{d'}=(i\sin n\phi\sin\theta,-i\sin n\phi\cos\theta,0)$. With this evolution operator $U(t,0)$
we can know the state at arbitrary time in periodic driving system.

\section{Examples}

\begin{figure}[htbp]
\centering
\includegraphics[scale=0.35]{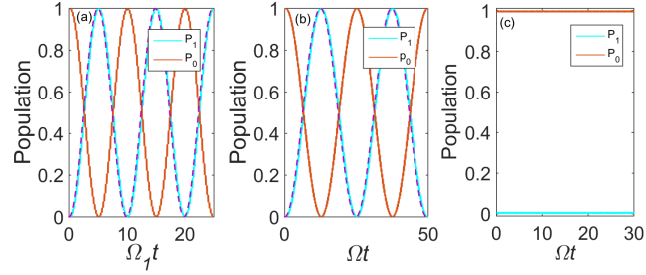}
\caption{The population as a function of time where $P_0$ ($P_1$) is the
population of state $|0\rangle$ ($|1\rangle$). (a) The intensity
modulation, $\Delta/\Omega_1=30$, $\Omega_2/\Omega_1=2$. (b) The
frequency modulation, $\Delta_1/\Omega=20$, $\Delta_2/\Omega=30$. (c) The constant Rabi frequency and detuning, $\Delta_1/\Omega=20$. The dash lines represent the function $|\sin\Lambda t|^2$. } \label{fig:02}
\end{figure}

As an example, we show how to achieve population inversion
$|0\rangle\rightarrow|1\rangle$ in a two-level atomic system to which a
classical field is applied. The system Hamiltonian reads
\begin{eqnarray}
H_0=\omega_0|1\rangle\langle1|+\Omega e^{-i\omega_lt}|0\rangle\langle1|+H.c,
\end{eqnarray}
where $\omega_0$ and $\omega_l$ are the atomic transition frequency and laser frequency, respectively.
$\Omega$ is the coupling constant. In the interaction picture, it becomes
$H=\Delta|1\rangle\langle1|+\Omega|0\rangle\langle1|+H.c.$, where
the detuning $\Delta=\omega_0-\omega_l$.
Comparing with Eq. (1) one obtains $\vec{d}=(\Omega,0,\frac{\Delta}{2})$.
Note that it is easy to control laser intensity to change the coupling
constant and regulate laser frequency to change the detuning. Here
we adopt two different ways to achieve population inversion, i.e.,
by manipulating the coupling constants $\Omega_1$ and $\Omega_2$
with a fixed large detuning $\Delta$ (called intensity
modulation) or by manipulating the detunings $\Delta_1$ and
$\Delta_2$ with a fixed coupling constant $\Omega$ (called
frequency modulation). Figs. \ref{fig:02}(a) and \ref{fig:02}(b)
demonstrate Rabi-like oscillations in dynamics evolution  by
periodic intensity and frequency modulations.
For comparison we plot the dynamics evolution without periodic
modulation in Fig. \ref{fig:02}(c).
Note that the frequency
of the Rabi-like oscillation  approximately equals to
$|\Lambda|$ (see the dash lines in Fig. \ref{fig:02}), which is in  good agreement with the effective
coupling strength of the effective Hamiltonian (4).

\begin{figure}[hb]
\centering
\includegraphics[scale=0.38]{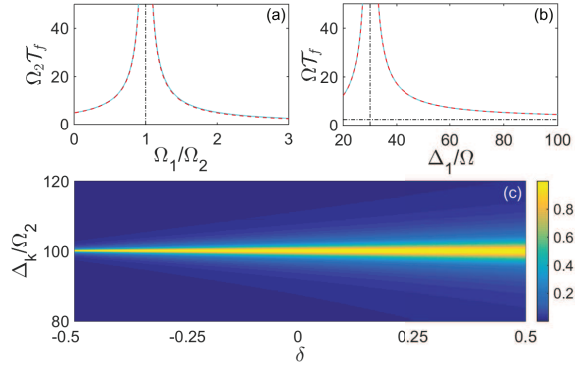}
\caption{The minimal time $\mathcal{T}_f$ as a function of (a) the coupling constant $\Omega_1$ in the
intensity modulation, $\Delta/\Omega_2=30$; (b) the detuning $\Delta_1$ in the
frequency modulation, $\Delta_2/\Omega=30$. The solid and dash
lines are the exact and approximate results, respectively. The
horizontal line (dot-dash line) is for
$\mathcal{T}_f'=\frac{\pi^2}{4\Omega}$. (c) The population with different detunings
$\Delta_k$ in the intensity modulation, where the square-well
pulse is described  by  $\Omega_1/\Omega_2=3$ and
$\Delta_1/\Omega_2=100$ for the transition
$|0\rangle\leftrightarrow|1\rangle$.  } \label{fig:03}
\end{figure}

The total time needed to achieve population inversion is,
\begin{eqnarray}   \label{5}
\mathcal{T}=\frac{(4m+1)\pi^2(|\vec{d_1}|+|\vec{d_2}|)}{4\phi|\vec{d_1}||\vec{d_2}|},~~m=0,1,2,...
\end{eqnarray}
As shown in Eq.(\ref{5}), the minimal time $\mathcal{T}_f$ ($m=0$) is determined by both the
coupling constants and the detunings. This gives rise to a question
whether there exists a set of parameters $\{\Omega_j, \Delta_j\}$ to
make $\mathcal{T}_f$ minimum. By expanding Eq.(\ref{5}) up to the
first order in $\phi$ (i.e., $\tan \phi\sim \phi$), it is not hard
to find that
$\mathcal{T}_f\simeq\frac{\pi^2}{4}\frac{\Delta_1+\Delta_2}{|\Delta_1\Omega_2-\Delta_2\Omega_1|}$.
Obviously, in the intensity modulation,
the minimal time $\mathcal{T}_f$
becomes short when the difference between two coupling constants is great.
When the coupling constant $\Omega_1$ approaches the coupling
constant $\Omega_2$, the minimal time $\mathcal{T}_f$
approaches infinity. This is not surprise since the system in this
situation returns to the case of constant driving with large
detuning, and the dynamics is frozen when $\Omega_1=\Omega_2$. In the
frequency modulation, the main results are similar to those
of the intensity modulation. The difference is that the minimal time
$\mathcal{T}_f$ almost remains unchanged when only one of the detunings varies, and it
asymptotically approaches $\frac{\pi^2}{4\Omega}$. Those
observations can be verified by numerical calculations, as shown in
Figs. \ref{fig:03}(a) and \ref{fig:03}(b). We can find  that the first-order
approximation is good enough to describe the total time
$\mathcal{T}$ in this case.

\section{Robustness  of the scheme and its extension  to three-level systems}

For a two-level system, the detuning (labelled as $\Delta_1$)
might be different from that used in the
calculation. One then asks whether this dismatch
affects population inversion in this scheme. In Fig.
\ref{fig:03}(c), we plot the population on the levels other than the
target, where the detunings of these levels to the level $|0\rangle$
are denoted by $\Delta_k$. We find  in Fig. \ref{fig:03}(c) that the
effect is remarkable.   E.g., the population is less than 0.1 when
the detunings have 5\% deviation from  the  detuning $\Delta_1$.
Fig. \ref{fig:03}(c) also shows that deviations in coupling constants
$\Omega_1$ [the deviation in the coupling constant is defined  as
$\Omega_1'=(1+\delta)\Omega_1$] has small  influence on the
population inversion $|0\rangle\rightarrow|1\rangle$.

\begin{figure}[htbp]
\centering
\includegraphics[scale=0.37]{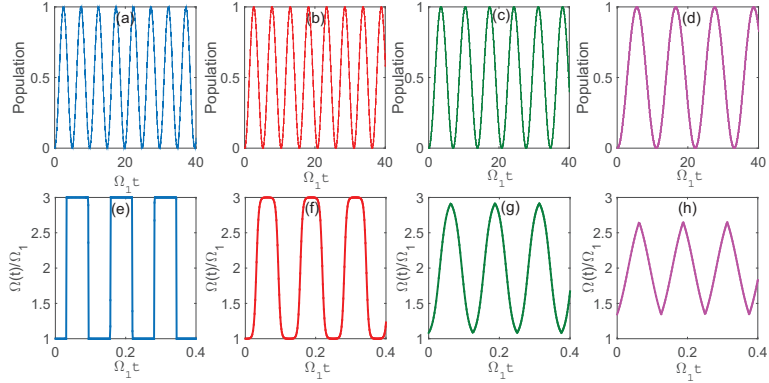}
\caption{The population $P_1$ as a function of time with different
$\gamma$ in the intensity modulation. (a) $\gamma=10000$. (b) $\gamma=300$. (c) $\gamma=100$.
(d) $\gamma=50$. The bottom panels (e)-(h) are the square-well field
corresponding to the top panels (a)-(d), respectively. The parameters of perfect
square-well field are $\Omega_2/\Omega_1=3$ and $\Delta/\Omega_1=50$.} \label{fig:s1}
\end{figure}

Note that it may be difficult to obtain a perfect square-well field in practice.
Thus we should analyze  how  deviations of the applied field from
the perfect square-well one  affect the population transfer. Taking the
intensity modulation as an example, we replace the perfect
square-well field with an approximate square-well field to study
this effect. Assume that the approximate square-well field takes
\begin{eqnarray}
\Omega(t)=\left \{
\begin{array}{ll}
    \Omega_2+\frac{\Omega_1-\Omega_2}{1+e^{\gamma(t-\frac{t_1}{2})}}, & t<0.5t_1,\\
    \Omega_2+\frac{\Omega_1-\Omega_2}{1+e^{-\gamma(t-T+\frac{t_1}{2})}}, & t\geq0.5t_1, \\
\end{array}
\right.
\end{eqnarray}
where $t_1=\frac{\pi}{2|\vec{d}_1|}$. We adopt a convention that
$t=(t$ mod $T)$ if $t>T$, and the parameter $\gamma>0$ is used to
adjust the hardness of the square-well field. Apparently, the larger
$\gamma$ is, the harder the square-well field would be. When
$\gamma\rightarrow+\infty$, this expression approaches to the
perfect square-well field. In Fig. \ref{fig:s1} we plot the
population $P_1$ when the square-well field is not perfect. For
small deformation of the square-well field (i.e., $\gamma$ is not
very small), it does not have  obvious influences  on the system
dynamics [cf. Figs. \ref{fig:s1}(a) and \ref{fig:s1}(b)].
Interestingly, even the square-well field approximately takes a
triangular shape, as shown in Fig. \ref{fig:s1}(d), population
transfer can also be realized, but the total time to reach
population inversion has been changed.

\begin{figure}[htbp]
\centering
\includegraphics[scale=0.35]{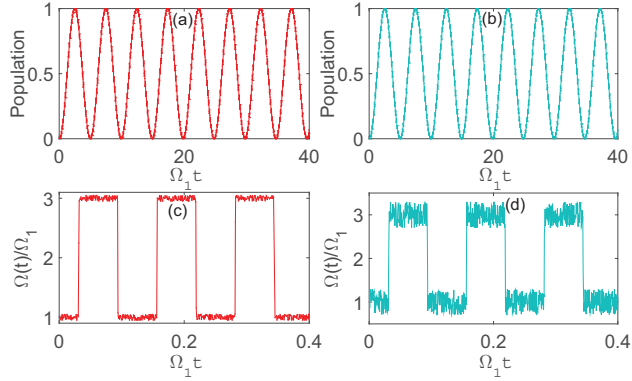}
\caption{The population $P_1$ as a function of time with noisy square-well field. (a) The noise $\epsilon(t)$ is generated
randomly in  the interval $[-0.05,0.05]$.  (b) The noise $\epsilon(t)$ is
generated randomly in the interval $[-0.3,0.3]$. The bottom panels
(c) and (d) are the noisy square-well fields corresponding to the top
panels (a) and (b), respectively. For perfect square-well fields, we employ $\Omega_2/\Omega_1=3$, and
$\Delta/\Omega_1=50$ in the intensity modulation.} \label{fig:s2}
\end{figure}

In actual modulation, there might exist  errors  such as
perturbations or noises in the coupling constant or the detuning.
Taking the intensity modulation as an example again, we add  noise
terms $\epsilon(t)$ into the perfect square-well field, which are
taken randomly in a certain interval. The results are plotted in
Fig. \ref{fig:s2}, which suggests  that the system dynamics is
robust against  noises.

\begin{figure}[htbp]
\centering
\includegraphics[scale=0.3]{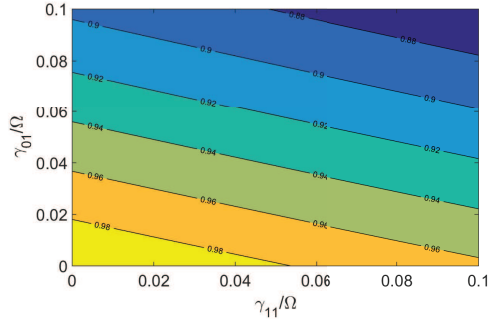}
\caption{ The population versus the dissipation rate $\gamma_{01}$
and the dephasing rate $\gamma_{11}$. The  initial state is
$|0\rangle$, $\Delta_1/\Omega=30$, $\Delta_2/\Omega=300$.}
\label{fig1a}
\end{figure}

So far, we do not consider the effect of environment yet. In the
presence of decoherence, we can use the following master equation to
describe the system dynamics \cite{breuer02},
\begin{eqnarray}
\dot{\rho}(t)=-i[H(t),\rho(t)]+\mathcal{L}_{01}(\rho)+\mathcal{L}_{11}(\rho),
\end{eqnarray}
where
$\mathcal{L}_{01}(\rho)=\frac{\gamma_{01}}{2}(2\sigma_{01}\rho\sigma_{10}-\sigma_{10}\sigma_{01}\rho
-\rho\sigma_{10}\sigma_{01})$ and
$\mathcal{L}_{11}(\rho)=\frac{\gamma_{11}}{2}(2\sigma_{11}\rho\sigma_{11}-\sigma_{11}\sigma_{11}\rho
-\rho\sigma_{11}\sigma_{11})$. Here, $\gamma_{01}$ and $\gamma_{11}$
are the dissipation  and the dephasing rate,  respectively.
$\sigma_{ij}=|i\rangle\langle j|$ for $i,j\in\{0,1\}$. Fig.
\ref{fig1a} shows the population of state $|1\rangle$ versus
the dissipation and dephasing rates. We find  that the population transfer
suffers  from the dissipation rate much more than the dephasing rate.

\begin{figure}[htbp]
\centering
\includegraphics[scale=0.35]{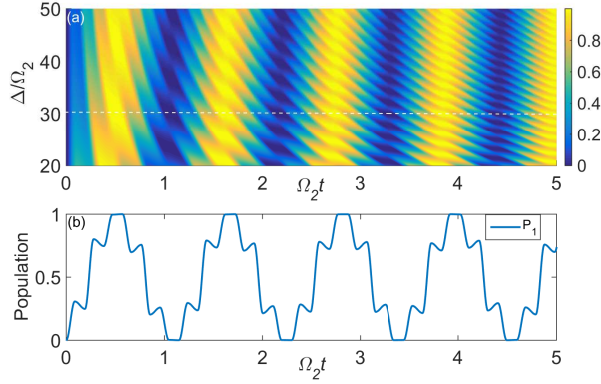}
\caption{(a) The population as a function of time and detuning,
$\Omega_1/\Omega_2=10$.  (b) The population as a function of
time when $\Delta/\Omega_2=30$ (the dash line), where the parameters
approximately satisfy Eq.(\ref{18}).  } \label{fig:04}
\end{figure}

Next we study the situation that the detuning is not very large.
That is, in the intensity manipulation,  one of coupling constants
(say, $\Omega_1$) is not very small compared to the detuning. If the
system parameters   satisfy one of the following equations,
\begin{eqnarray} \label{18}
d_{1x}\sin(m\phi)+d_{1z}\cos(m\phi)=|\vec{d_1}|,~~m=1,2,3,...,
\end{eqnarray}
the system  would stay in the target state for a long  time with
respect to the case without this condition, as shown in Fig.
\ref{fig:04}(b). One can observe plateaus  in the time evolution when
the population reaches one. Those plateaus provide us with large
opportunity to stop the population transfer, and make the system
stay at the high level forever. Physically this originates from the
small coupling constant (such as $\Omega_2$) while another is very
large. In this case, the system is shortly  frozen when it arrives
at one of the two levels.

This technique  can be extended to achieve an arbitrary coherent
superposition state, for example,
$|\Psi_T\rangle=\cos\Theta|0\rangle+\sin\Theta|1\rangle$ in the
$\Lambda$-type (or $V$-type) three-level system, where two lower
levels couple to another through  classical fields with coupling
constant $\Omega_1$ and $\Omega_2$, respectively. Both detunings are
assumed to be the same $\Delta$. At first, we should choose the
ratio of $\Omega_1/\Omega_2$ to equal $\tan(\Theta)$ given in
$|\Psi_T\rangle$. By periodically changing the detunings $\Delta_a$
and $\Delta_b$ with fixed coupling constant $\Omega_1$ and
$\Omega_2$, one can achieve the state $|\Psi_T\rangle$. If the
three-level atom is placed in a cavity, and one of the transition is
driven by the cavity field while the other is driven by a classical field, we
can realize single photon storing by manipulating  the frequency and
the intensity of the classical field (for details, see Appendix).

\begin{figure}[htbp]
\centering
\includegraphics[scale=0.35]{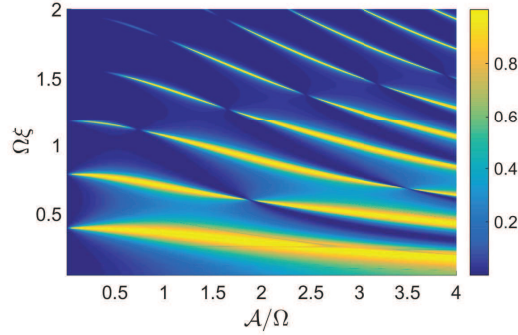}
\caption{The population versus   $\{\mathcal{A}, \xi\}$ of the
Gaussian pulse, $\Delta/\Omega=2$. } \label{fig:04a}
\end{figure}

In experiments, the coupling might not be  a constant for  laser
pulse drivings. Further investigation shows that the scheme works
well provided that the pulse  takes a Gaussian form
$\Omega(t)=\mathcal{A}e^{-\frac{(t-4\xi)^2}{2\xi^2}}$. Here
$\mathcal{A}$ and $\xi$ denote  the amplitude and the width of the
Gaussian pulse, respectively. In this case the period $T$ is the
time width of full Gaussian pulse, and we truncate the width of
Gaussian pulse with $T=8\xi$ in the following. The evolution
operator can be written as
$U(T,0)=\mathbb{T}e^{-i\int_{0}^{T}H(t')dt'}$ with $\mathbb{T}$
denoting  the time-ordering operator. It is difficult to  find an
analytical expression for the evolution operator. Fortunately, we
can prove  that after $n$ Gaussian pulses the system evolution
operator takes following form
\begin{eqnarray}  \footnotesize
U(nT,0)=\left(
\begin{array}{cc}
\cos(n\vartheta)-i\frac{Q'\sin(n\vartheta)}
{\sqrt{Q'^2+R'^2}}  & -\frac{R'}{\sqrt{Q'^2+R'^2}}\sin(n\vartheta)e^{i\theta'} \\
\frac{R'}{\sqrt{Q'^2+R'^2}}\sin(n\vartheta)e^{-i\theta'}
& \cos(n\vartheta)+i\frac{Q'\sin(n\vartheta)}{\sqrt{Q'^2+R'^2}} \\
\end{array}
\right).   \nonumber
\end{eqnarray}
where $\cos\vartheta=P'$ and the constants $\{P', Q', R', \theta'\}$
are jointly determined by $\mathcal{A}$, $\xi$, and $\Delta$. One easily
observes that if $Q'\neq0$, namely there exists an imaginary part in
the evolution operator $U(T,0)$, the population inversion cannot be
perfectly obtained. We can adjust the parameters $\{\mathcal{A},
\xi, \Delta\}$   to make $Q'$ vanish and approximately realize
population inverse after $N$ Gaussian pulses with
$N=\frac{(4m+1)\pi}{2\vartheta},m=0,1,2,...$.  Fig. \ref{fig:04a}
demonstrates how the parameters  $\{\mathcal{A}, \xi, \Delta\}$ of
the Gaussian pulse affects  the system dynamics. It indicates that
we can still achieve population inversion with appropriate Gaussian
pulses even though the system is driven by lasers off-resonant with
the transition energy.

\section{Extension to Rabi model}

The Rabi model
\cite{rabi36,rabi37,solano11,albert12,hwang15} describes  a
two-level atom (system) coupled  to a single-mode electromagnetic
field, which  provides us with a simplest   example for
light-matter interactions. The Hamiltonian describing  such a system
reads
\begin{eqnarray} \label{12}
H=\omega_b a^{\dag}a+\frac{\omega_0}{2}\sigma_z+\Omega
\sigma_x(a^{\dag}+a),
\end{eqnarray}
where $\sigma_x$ and $\sigma_z$ are Pauli matrices for the two-level
system with  level splitting $\omega_0$. $a$ ($a^{\dag}$) is the
annihilation (creation) operator of the single mode field with
frequency $\omega_b$. $\Omega$ denotes the coupling constant between
the two-level system and the field.
Note that the Hamiltonian (\ref{12})
describes  the simplest interaction between atom and field,
and can be solved analytically \cite{braak11,chen12}.
Rewriting the Hamiltonian (\ref{12}) in the Hilbert space spanned by
$\{|g,0\rangle, |e,0\rangle, |g,1\rangle,
|e,1\rangle,...,|g,n\rangle, |e,n\rangle,...\}$, where $|f,n\rangle$
denotes the state that the two-level system  in state $|f\rangle$
($f=e,g$) with  $n$ photons in the field, we have
\begin{eqnarray}
H=\left(
  \begin{array}{cccccccc}
    -\frac{\omega_0}{2} & 0 & 0 & \Omega & 0 & 0 & 0 & \cdots \\
    0 & \frac{\omega_0}{2} & \Omega & 0 & 0 & 0 & 0 & \cdots \\
    0 & \Omega & -\frac{\omega_0}{2}+\omega_b & 0 & 0 & \Omega & 0 & \cdots \\
    \Omega & 0 & 0 & \frac{\omega_0}{2}+\omega_b & \Omega & 0 & 0 & \cdots \\
    0 & 0 & 0 & \Omega & -\frac{\omega_0}{2}+2\omega_b & 0 & 0 & \cdots \\
    0 & 0 & \Omega & 0 & 0 & \frac{\omega_0}{2}+2\omega_b & \Omega & \cdots \\
    0 & 0 & 0 & 0 & 0 & \Omega & -\frac{\omega_0}{2}+3\omega_b & \cdots \\
    \vdots & \vdots & \vdots & \vdots & \vdots & \vdots & \vdots & \ddots \\
  \end{array}
\right).
\end{eqnarray}
Then the system dynamics is governed  by Liouville equation
\begin{eqnarray}
\dot{\rho}(t)=-i[H,\rho(t)],
\end{eqnarray}
where $\rho(t)$ is the density operator for the composite
system (atom plus field).

\begin{figure}[htbp]
\centering
\includegraphics[scale=0.4]{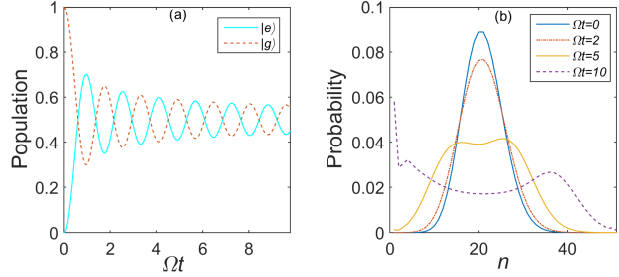}
\caption{(a) The population of $|e\rangle$ and $|g\rangle$ as a
function of time with resonant atom-field couplings (i.e.,
$\Delta=\omega_0-\omega_b=0$). (b) The probability that there are $n$ photons
in the field  at different time. The field is initially in a
coherent state given by Eq.(\ref{13}) with the average photon number $\langle n\rangle=20$.
The photon number  is truncated at 51 in numerical calculations.} \label{fig:09}
\end{figure}

Suppose the two-level system is in the ground state
$\rho^{s}(0)=|g\rangle\langle g|$ and the field  is in a coherent
state initially, namely
\begin{eqnarray}   \label{13}
\rho^{b}_{nn}(0)=\frac{\langle n\rangle^{n}e^{-\langle
n\rangle}}{n!}.
\end{eqnarray}
We find that it is still unavailable to
realize population inversion  perfectly starting from
$\rho(0)=\rho^{s}(0)\otimes\rho^{b}(0)$ by the constant driving field, even though the transition frequency of the
two-level system is on resonance with the frequency of field (i.e.,
$\Delta=\omega_0-\omega_b=0$), as shown in Fig. \ref{fig:09}(a). This is quite different
from the Jaynes-Cummings (JC) model \cite{jaynes63}
where it is applied  the rotation-wave
approximation to this model. Physically, this result  originates from the
emergence of counter-rotating part, i.e., $\sigma^{+}a^{\dag}$ and
$\sigma^{-}a$ in Hamiltonian (\ref{12}), where $\sigma^{+}$
($\sigma^{-}$) is the raising (lowering) operator for the two-level
system. From the aspect of the field, it does not remain
coherent state any more, as shown in Fig. \ref{fig:09}(b).

\begin{figure}[htbp]
\centering
\includegraphics[scale=0.4]{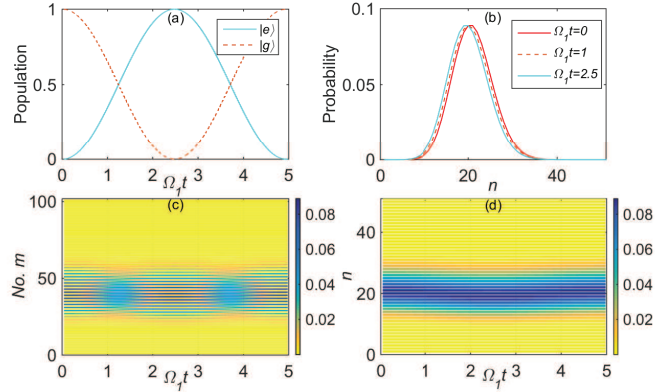}
\caption{(a) The population of $|e\rangle$ and $|g\rangle$ as  a
function of time in the intensity modulation, $\Delta/\Omega_1=300$,
$\Omega_2/\Omega_1=3$. (b) The probability that there are $n$ photons
in the field at different time. The field is initially in a
coherent state described by Eq.(\ref{13}) with $\langle
n\rangle=20$. (c) The occupation of No. $m$ basis as a function of
time, where the basis are ordered as $\{|g,0\rangle, |e,0\rangle,
|g,1\rangle, |e,1\rangle,..., |g,50\rangle, |e,50\rangle,
|g,51\rangle\}$. (d) The probability of $n$ photons in the field  as
a function of time.} \label{fig:10}
\end{figure}

\begin{figure}[htbp]
\centering
\includegraphics[scale=0.4]{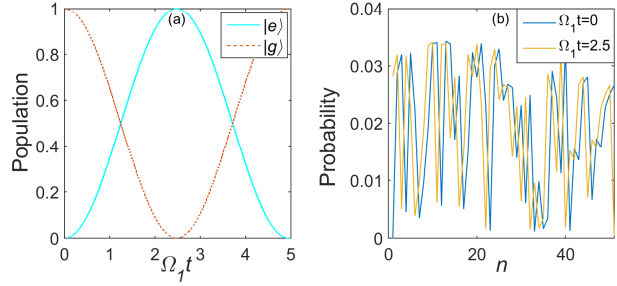}
\caption{(a) The population at $|e\rangle$ and $|g\rangle$ as  a
function of time in the intensity modulation. (b) The probability
that there are $n$ photons in the field at different time. The field
is in random state initially.} \label{fig:11}
\end{figure}

In contrast, by the intensity modulation, we can realize the
population transfer as shown in the following. Keeping  the
detuning ($\Delta=\omega_b-\omega_0$) fixed, and modulating the
coupling constant $\Omega_1$ and $\Omega_2$,  we  plot the evolution
of the population  in  Fig. \ref{fig:10}(a). We find that
population transfer  from $|g\rangle$ to $|e\rangle$ occurs at
$\Omega_1t=2.5$. An inspection of Fig. \ref{fig:10}(b) shows that the  state
of the field (the coherent state) remains unchanged in the dynamics
[see Fig. \ref{fig:10}(d) as well]. Furthermore, we find that
population transfer  can occur for a wide range of initial states of
the  field, as shown in Fig. \ref{fig:11}, where the initial state of the
field is $\rho^{b}_{nn}(0)=\mathcal{P}_n$. Here $\mathcal{P}_n$ are
a set of random numbers  in  interval $[0,1]$ and satisfy the
normalization condition $\sum_{n}\mathcal{P}_n=1$.

\section{Discussion and conclusion}

On the other hand, this method can  be applied to study Landau-Zener-St\"{u}ckelberg
interference \cite{shevchenko10,huang11,ferron12,cao13,dupont13},
which has been observed in a superconducting qubit under periodic
modulation
\cite{izmalkov04,oliver05,sillanpaa06,izmalkov08,lahaye09,johansson09,silveri15}.
Obviously, the Landau-Zener transition can also be fore-sight by
using our method. In addition, the scheme can be used to realize
population inversion for Rydberg states, e.g., from the ground state
$|g\rangle\equiv5S_{1/2}|F=2,m=2\rangle$ to the Rydberg state
$|r\rangle\equiv60S_{1/2}$ of the cold $^{87}$Rb atom. Since this
atomic transition frequency is in the ultraviolet region, it
conventionally employs 780-nm and 480-nm lasers via the intermediate
state $|e\rangle\equiv5P_{3/2}|F=3,m=3\rangle$ under two-photon
resonance condition \cite{thaichroen15}. In our scheme, only 780-nm
lasers can work. The detuning between atomic transition frequency
and 780-nm laser frequency is about $\Delta=6.25\times10^{8}$ MHz,
which is in the far-off-resonant regime. The two  coupling strengths
we employ are $\Omega_1=2\pi\times30$ MHz and
$\Omega_2=2\pi\times300$ MHz. We can set $t_1\simeq t_2\simeq5.03$
ps, and the  addressing period $T\simeq10.06$ ps. {Note that the period of driving field should be
modulated with high precision in the far-off-resonant regime.} As a result the total
time of realizing this transition is about 2.9 $\mu$s.

The other possible application is to achieve long-lived excited
nuclear states. The transition energy between the nuclear states is
much larger  than the   energy of  X-ray lasers. To compensate the
energy difference between nuclear transition and X-ray laser, one
envisages accelerated nuclei interacting with x-ray laser pulses. In
the strong acceleration regime, the resonance condition cannot be
satisfied very well for requiring high degree of accuracy of
relativistic factor \cite{burvenich06,liao11}. As a result it
significantly influences population transfer. Our calculations
shows that this problem can be effectively solved by a series of
Gaussian pulses. Take the E1 transition in $^{223}$Ra as an example.
It is impossible to transfer the population to excited state with
only a single pulse when the detuning $\Delta$ is about $0.1$ eV
\cite{burvenich06}. However, if we take a laser intensity
$I=9\times10^{22}$ W/cm$^2$ and the width of Gaussian pulse is 31
fs, the   transition can be obtained by 5 Gaussian pulses.

In conclusion, a scheme to realize population transfer by
far-off-resonant drivings is proposed. By two sequentially applied
lasers with different frequency or intensity, population transfer
can be perfectly achieved in two-level systems. This proposal can be
extended to $N$-level systems and to the case with cavity fields
instead of classical fields. Furthermore, it can be exploited to
achieve population transfer  in the Rabi model regardless the form
of bosonic field, which has been experimentally explored in the
photonic analog simulator \cite{crespi12}.

The scheme can be applied to population transfer from ground state
to Rydberg states or Rydberg state preparation. In X-ray quantum
optics \cite{adams13}, it may also find applications, since the
lack of $\gamma$-ray lasers and huge transition energy inside the
nuclei. This scheme can work with both square-well pulse and
Gaussian pulse, and it is robust against dissipation and dephasing
in the sense that the life-time of Rydberg states as well as the
nuclear excited state is long. By manipulating the system
parameters, we can make the system stay at the excited level longer.
This work paves  a new avenue in preparation  and manipulation of
quantum states with off-resonant driving fields, and might find
potential applications in the field of quantum optics.

\section*{Appendix: Three-level system  in  periodic square-well fields}

\begin{figure}[htbp]
\centering
\includegraphics[scale=0.25]{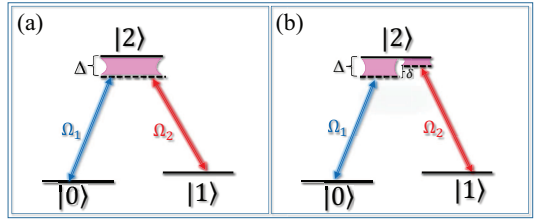}
\caption{The level configuration of a three-level system.}
\label{fig:01}
\end{figure}

We now demonstrate how to apply the periodic modulation to the other
structure of quantum systems--three-level system. As shown in Fig.
\ref{fig:01}(a), we consider a $\Lambda$-type three-level system (it also
works in $V$-type system) that two levels coupled to the third  with
coupling constants $\Omega_1$ and $\Omega_2$, respectively. $\Delta$
denotes the  detuning in this system. Without loss of generality, we
assume $\Omega_1$ and $\Omega_2$ to be real. In the interaction
picture, the Hamiltonian in the Hilbert space spanned by
$\{|0\rangle, |1\rangle, |2\rangle\}$ reads
\begin{eqnarray}
H=\left(
           \begin{array}{ccc}
             0 & 0 & \Omega_1 \\
             0 & 0 & \Omega_2 \\
             \Omega_1 & \Omega_2 & \Delta \\
           \end{array}
         \right).
\end{eqnarray}
The evolution operator becomes
\begin{eqnarray}
U(t,0)=e^{-iHt}=e^{-\frac{i\Delta t}{2}}\left(
                                   \begin{array}{ccc}
                                     B_1 & B_2 & B_4 \\
                                     B_2 & B_3 & B_5 \\
                                     B_4 & B_5 & B_6 \\
                                   \end{array}
                                 \right),
\end{eqnarray}
where
$B_1=\frac{\Omega_2^2}{\Omega_1^2+\Omega_2^2}e^{\frac{i}{2}\Delta t}
+\frac{\Omega_1^2}{\Omega_1^2+\Omega_2^2}(\cos\frac{yt}{2}+
i\frac{\Delta}{y}\sin\frac{yt}{2})$,
$B_2=-\frac{\Omega_1\Omega_2}{\Omega_1^2+\Omega_2^2}e^{\frac{i}{2}\Delta
t} +\frac{\Omega_1\Omega_2}{\Omega_1^2+\Omega_2^2}(\cos\frac{yt}{2}+
i\frac{\Delta}{y}\sin\frac{yt}{2})$,
$B_3=\frac{\Omega_1^2}{\Omega_1^2+\Omega_2^2}e^{\frac{i}{2}\Delta t}
+\frac{\Omega_2^2}{\Omega_1^2+\Omega_2^2}(\cos\frac{yt}{2}+
i\frac{\Delta}{y}\sin\frac{yt}{2})$, $B_4=-2 i
\frac{\Omega_1}{y}\sin\frac{yt}{2}$, $B_5=-2 i
\frac{\Omega_2}{y}\sin\frac{yt}{2}$,
$B_6=\cos\frac{yt}{2}-i\frac{\Delta}{y}\sin\frac{yt}{2}$, and
$y=\sqrt{4\Omega_1^2+4\Omega_2^2+\Delta^2}$. When the initial state
is $|2\rangle$ and in the large detuning limit (i.e.,
$\Delta\gg\Omega_1,\Omega_2$), we find that $B_4\simeq B_5\simeq0$
in the evolution operator. Thus the population transfer can not be
realized in this case.

Now we turn to study the system dynamics under periodic frequency
modulation. That is, the Hamiltonian is $H_a$ in the time interval
$[0,t_1]$ and  $H_b$ in the time interval $(t_1,T]$, where
$H_j=\left(
           \begin{array}{ccc}
             0 & 0 & \Omega_1 \\
             0 & 0 & \Omega_2 \\
             \Omega_1 & \Omega_2 & \Delta_j \\
           \end{array}
         \right), j=a,b.$
Note that both $\Delta_a$ and $\Delta_b$ are chosen to be very
large, so the system dynamics is frozen with only a  single driving.
The evolution operator of one period can be calculated as
\begin{eqnarray}
U(T,0)=e^{-iH_bt_2}e^{-iH_at_1}
          =\frac{e^{-\frac{i}{2}(\Delta_at_1+\Delta_bt_2)}}{y_ay_b}\left(
                                   \begin{array}{ccc}
                                     B_1' & B_2' & B_4' \\
                                     B_2' & B_3' & B_5' \\
                                     -B_4' & -B_5' & B_6' \\
                                   \end{array}
                                 \right),
\end{eqnarray}
where
$B_1'=\frac{1}{\Omega_1^2+\Omega_2^2}[y_ay_b\Omega_2^2e^{\frac{i}{2}(\Delta_at_1+\Delta_bt_2)}
-\Omega_1^2(4\Omega_1^2+4\Omega_2^2+\Delta_a\Delta_b)]$,
$B_2'=-\frac{1}{\Omega_1^2+\Omega_2^2}[y_ay_b\Omega_1\Omega_2e^{\frac{i}{2}(\Delta_at_1+\Delta_bt_2)}+
\Omega_1\Omega_2(4\Omega_1^2+4\Omega_2^2+\Delta_a\Delta_b)]$,
$B_3'=\frac{1}{\Omega_1^2+\Omega_2^2}[y_ay_b\Omega_1^2e^{\frac{i}{2}(\Delta_at_1+\Delta_bt_2)}
-\Omega_2^2(4\Omega_1^2+4\Omega_2^2+\Delta_a\Delta_b)]$,
$B_4'=2\Omega_1(\Delta_b-\Delta_a)$,
$B_5'=2\Omega_2(\Delta_b-\Delta_a)$, and
$B_6'=-4(\Omega_1^2+\Omega_2^2)-\Delta_a\Delta_b$. We have set
$y_at_1=y_bt_2=\pi$, $y_j=\sqrt{4\Omega_1^2+4\Omega_2^2+\Delta_j^2},
j=a,b$. Since the evolution operator is time-dependent, it is
difficult to calculate the system dynamics analytically for $n$
evolution periods. Further observations demonstrate that if the
system state
$|\Psi(t)\rangle=c_0(t)|0\rangle+c_1(t)|1\rangle+c_2(t)|2\rangle$
satisfies the condition
$\frac{c_0(t)}{c_1(t)}=\frac{\Omega_1}{\Omega_2}$, the evolution
operator would be time-independent. Then the evolution operator
after $n$ evolution periods becomes
\begin{eqnarray}   \footnotesize
U(t,0)=U(nT,0)=\frac{e^{-\frac{in}{2}(\Delta_at_1+\Delta_bt_2)}}{(y_ay_b)^n}
\left(
                  \begin{array}{ccc}
                    \frac{\Omega_1^2}{2(\Omega_1^2+\Omega_2^2)}(s_1^n+s_2^n) & \frac{\Omega_1\Omega_2}{2(\Omega_1^2+\Omega_2^2)}(s_1^n+s_2^n) &
                    \frac{i\Omega_1}{2\sqrt{\Omega_1^2+\Omega_2^2}}(s_1^n-s_2^n) \\
                    \frac{\Omega_1\Omega_2}{2(\Omega_1^2+\Omega_2^2)}(s_1^n+s_2^n) & \frac{\Omega_2^2}{2(\Omega_1^2+\Omega_2^2)}(s_1^n+s_2^n) &
                    \frac{i\Omega_2}{2\sqrt{\Omega_1^2+\Omega_2^2}}(s_1^n-s_2^n) \\
                    \frac{-i\Omega_1}{2\sqrt{\Omega_1^2+\Omega_2^2}}(s_1^n-s_2^n) & \frac{-i\Omega_2}{2\sqrt{\Omega_1^2+\Omega_2^2}}(s_1^n-s_2^n) &
                    \frac{1}{2}(s_1^n+s_2^n) \\
                  \end{array}
                \right),   \nonumber
\end{eqnarray}
where $s_1=d_1+id_2=|s|e^{i\varphi}$,
$s_2=d_1-id_2=|s|e^{-i\varphi}$, $\tan\varphi=\frac{d_2}{d_1}$,
$|s|=\sqrt{d_1^2+d_2^2}$,
$d_1=-4\Omega_1^2-4\Omega_2^2-\Delta_a\Delta_b$, and
$d_2=2(\Delta_a-\Delta_b)\sqrt{\Omega_1^2+\Omega_2^2}$. When the
initial state is $|\Psi(0)\rangle=|2\rangle$, the final state
becomes
\begin{eqnarray}   \label{11}
|\Psi(t)\rangle=U(nT,0)|\Psi(0)\rangle=\mathcal{N}\left(
                                            \begin{array}{c}
                                              \Omega_1\sin(n\varphi) \\
                                              \Omega_2\sin(n\varphi) \\
                                              -i\sqrt{\Omega_1^2+\Omega_2^2}\cos(n\varphi) \\
                                            \end{array}
                                          \right),
\end{eqnarray}
where $\mathcal{N}$ is a normalization constant. Note that
whether the system evolves or not depends only on the phase
$\varphi$, regardless of  the large detuning condition, and the
Rabi-like oscillation frequency   is determined by the phase
$\varphi$.

\begin{figure}[htbp]
\centering
\includegraphics[scale=0.45]{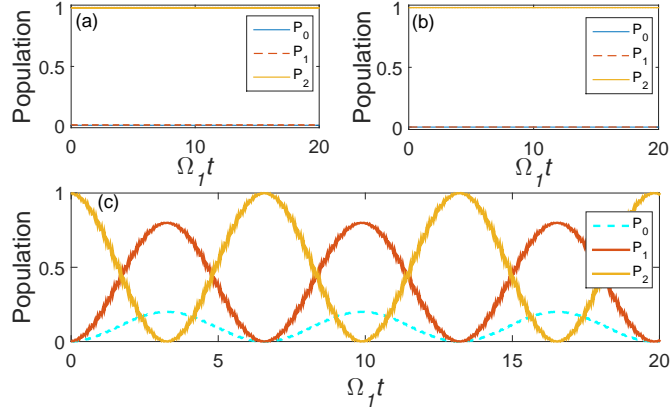}
\caption{The population as a function of time where the initial state is
$|2\rangle$. (a) $\Delta/\Omega_1=50$, $\Omega_2/\Omega_1=2$. (b)
$\Delta/\Omega_1=100$, $\Omega_2/\Omega_1=2$. (c) $\Delta_a/\Omega_1=50$,
$\Delta_b/\Omega_1=100$, $\Omega_2/\Omega_1=2$. } \label{fig:05}
\end{figure}

We next demonstrate how to realize an arbitrary coherent
superposition state
$|\Psi_T\rangle=\cos\Theta|0\rangle+\sin\Theta|1\rangle$ in the
$\Lambda$-type (or $V$-type) three-level system. We regulate the
detunings $\Delta_a$ and $\Delta_b$ and fix the coupling constants
$\Omega_1$ and $\Omega_2$. Especially, the ratio of
$\Omega_1/\Omega_2$ is modulated to equal $\tan\Theta$ in
$|\Psi_T\rangle$. Again, we study in the  large detuning regime.
According to Eq.(\ref{11}), if we set $N\varphi=\frac{\pi}{2}$, the
system would arrive at the coherent superposition state
$|\Psi_T\rangle$ and the total time of achieving $|\Psi_T\rangle$ is
about
\begin{eqnarray}
\mathcal{T}=\frac{\pi}{2\tan^{-1}|\frac{2(\Delta_a-\Delta_b)\sqrt{\Omega_1^2+\Omega_2^2}}
{4\Omega_1^2+4\Omega_2^2+\Delta_a\Delta_b}|}(\frac{\pi}{\sqrt{4\Omega_1^2+4\Omega_2^2+\Delta_a^2}}
+\frac{\pi}{\sqrt{4\Omega_1^2+4\Omega_2^2+\Delta_b^2}}).
\end{eqnarray}
For comparison, we plot the system dynamics under constant driving field with the detuning
$\Delta/\Omega_1=50$ in Fig. \ref{fig:05}(a) and $\Delta/\Omega_1=100$ in Fig.
\ref{fig:05}(b), where the fixed coupling constants satisfy $\Omega_2/\Omega_1=2$. One can find that  the system
dynamics is frozen due to the large detuning condition. However, as shown in
Fig. \ref{fig:05}(c), it is completely different in the frequency
modulation, where two tunable detunings are also $\Delta_a/\Omega_1=50$ and
$\Delta_b/\Omega_1=100$. Namely, the population transfer can be realized in
the later case.

Note that our theory can also be generalized  into   $N$-level
systems, where one level couples to the reminder levels with the
corresponding coupling constants $\Omega_1$, $\Omega_2$,...,
$\Omega_{N}$, and all the detunings are $\Delta$. By a similar
procedure shown above in the frequency modulation, we can obtain the
following coherent superposition state,
\begin{eqnarray}
|\Psi_T\rangle=\mathcal{N}(\Omega_1|0\rangle+\Omega_2|1\rangle+\cdots+\Omega_{N}|N-1\rangle),
\end{eqnarray}
where the coefficients $\Omega_1$, $\Omega_2$,..., $\Omega_{N}$ are
determined by coupling constants and $\mathcal{N}$ is a
normalization constant.

\begin{figure}[htbp]
\centering
\includegraphics[scale=0.4]{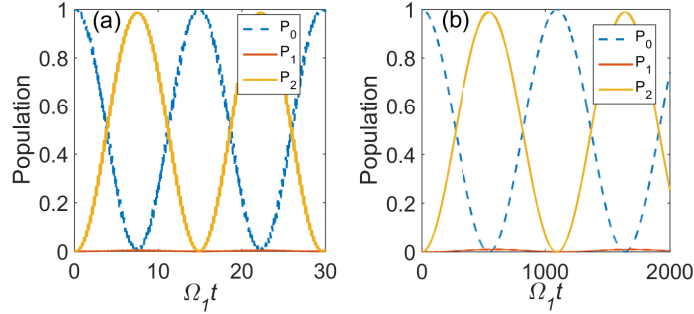}
\caption{The population as a function of time in frequency
modulation. The initial state is $|0\rangle$. (a)
$\Delta_{\alpha}/\Omega_1=40$, $\Delta_{\beta}/\Omega_1=20$, $\delta/\Omega_1=10$,
$\Omega_2=\Omega_1$. (b) $\delta_\alpha=0$, $\delta_\beta/\Omega_1=30$,
$\Delta/\Omega_1=10$, $\Omega_2/\Omega_1=\Omega_1$. } \label{fig:06}
\end{figure}

Next, we show how to modulate the detunings $\Delta$ or $\delta$ to
realize population transfer $|0\rangle\rightarrow|2\rangle$  in Fig.
\ref{fig:01}(b). Numerical simulation results are presented in Fig.
\ref{fig:06}. We regulate two detunings $\Delta_{\alpha}$ and
$\Delta_{\beta}$ in Fig. \ref{fig:06}(a), where two Hamiltonians
read $H_j=\left(
           \begin{array}{ccc}
             0 & 0 & \Omega_1 \\
             0 & \delta & \Omega_2 \\
             \Omega_1 & \Omega_2 & \Delta_j \\
           \end{array}
         \right), j=\alpha,\beta.$
Fig. \ref{fig:06}(b) demonstrates the dynamics by regulating two
detunings $\delta_\alpha$ and $\delta_\beta$, where two Hamiltonians
read $H_j=\left(
           \begin{array}{ccc}
             0 & 0 & \Omega_1 \\
             0 & \delta_j & \Omega_2 \\
             \Omega_1 & \Omega_2 & \Delta \\
           \end{array}
         \right), j=\alpha,\beta.$
It is shown that the system evolves with time in both modulations,
and one can obtain population transfer
$|0\rangle\rightarrow|2\rangle$ by choosing  the evolution time
$\mathcal{T}$.

\begin{figure}[htbp]
\centering
\includegraphics[scale=0.4]{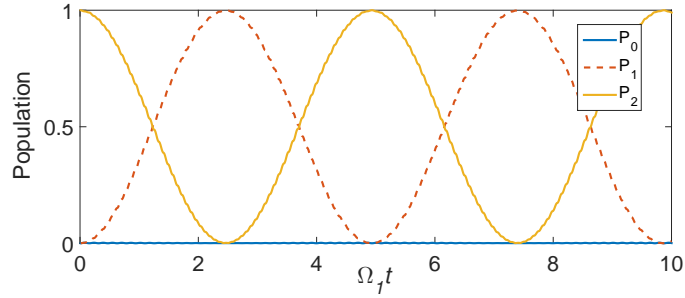}
\caption{The population as a function of time in intensity
modulation. The initial state is $|2\rangle$, $\Delta/\Omega_1=40$,
$\delta=0$, $\Omega_2=\Omega_1$, $\Omega_2'/\Omega_1=-1.$} \label{fig5}
\end{figure}

We should emphasize that the  population transfer
$|2\rangle\rightarrow|1\rangle$  can also be realized by intensity
modulation. That is, the Hamiltonian is $H_1'=\left(
           \begin{array}{ccc}
             0 & 0 & \Omega_1 \\
             0 & \delta & \Omega_2 \\
             \Omega_1 & \Omega_2 & \Delta \\
           \end{array}
         \right)$ in the time interval $[0,t_1]$, while the Hamiltonian is $H_2'=\left(
           \begin{array}{ccc}
             0 & 0 & \Omega_1 \\
             0 & \delta & \Omega_2' \\
             \Omega_1 & \Omega_2' & \Delta \\
           \end{array}
         \right)$ in the
time interval $(t_1,T]$. The results are plotted in Fig. \ref{fig5}.
If we put the three-level atom in a cavity, the transition  between
$|0\rangle\leftrightarrow|2\rangle$ is coupled by the cavity field
with the detuning $\Delta$, and the transition between
$|1\rangle\leftrightarrow|2\rangle$ is coupled by classical field
with the detuning $\Delta-\delta$. The Hilbert space of the
atom-cavity system is spanned by
$\{|0,1\rangle,|2,0\rangle,|1,0\rangle\}$, where $|m,n\rangle$
denotes the atom state $|m\rangle$ and $n$ photons in the cavity.
For the initial state $|0,1\rangle$, by combining the frequency and
intensity modulation of classical field to reach the state
$|1,0\rangle$, one can use it to realize single photon storing or
releasing.

\section*{Acknowledgments}

We thank Y. M. Liu for helpful discussions.  This work is supported
by the National Natural Science Foundation of China (NSFC) under Grants
No.11534002 and No.61475033.


\begin{thebibliography}{99}


\bibitem{scully97} M. O. Scully and M. S. Zubairy, \emph{Quantum Optics}
(Cambridge University, 1997).
\bibitem{nielsen10} M. Nielsen and I. Chuang, \emph{Quantum Computation and Quantum
Information} (Cambridge University, 2010).

\bibitem{saffman10} M. Saffman, T. G. Walker, and K. M{\o}lmer, ``Quantum information with Rydberg atoms,'' Rev. Mod. Phys. \textbf{82}, 2313 (2010).


\bibitem{schempp15} H. Schempp, G. G\"{u}nter, S. W\"{u}ster, M. Weidem\"{u}ller, and S. Whitlock, ``Correlated exciton transport in Rydberg-dressed-atom spin chains,'' Phys. Rev. Lett. \textbf{115}, 093002 (2015).
\bibitem{barredo15} D. Barredo, H. Labuhn, S. Ravets, T. Lahaye, A. Browaeys, and C. S. Adams, ``Coherent excitation transfer in a spin chain of three Rydberg atoms,'' Phys. Rev. Lett. \textbf{114}, 113002 (2015).


\bibitem{walker99} P. Walker and G. Dracoulis, ``Energy traps in atomic nuclei,'' Nature \textbf{399}, 35--40 (1999).
\bibitem{ledingham03} K. W. D. Ledingham, P. McKenna, and R. P. Singhal, ``Applications for nuclear phenomena generated by ultra-intense lasers,'' Science \textbf{300}, 1107--1111 (2003).
\bibitem{aprahamian05} A. Aprahamian and Y. Sun, ``Nuclear physics: Long live isomer research,'' Nature Physics \textbf{1}, 81--82 (2005).
\bibitem{paffy07} A. P\'{a}lffy, J. Evers, and C. H. Keitel, ``Isomer triggering via nuclear excitation by electron capture,'' Phys. Rev. Lett. \textbf{99}, 172502 (2007).
\bibitem{bergman98} K. Bergmann, H. Theuer, and B. W. Shore, ``Coherent population transfer among quantum states of atoms and molecules,'' Rev. Mod. Phys. \textbf{70}, 1003 (1998).
\bibitem{burvenich06} T. J. B\"{u}rvenich, J. Evers, and C. H. Keitel, ``Nuclear quantum optics with X-ray laser pulses,'' Phys. Rev. Lett. \textbf{96}, 142501 (2006).
\bibitem{altarelli06} M. Altarelli, R. Brinkmann, M. Chergui, W. Decking, B. Dobson, S. D\"{u}sterer, G. Gr\"{u}bel, W. Graeff, H. Graafsma, J. Hajdu, J. Marangos, J. Pfl\"{u}ger, H. Redlin, D. Riley, I. Robinson, J. Rossbach, A. Schwarz, K. Tiedtke, T. Tschentscher, I. Vartaniants, H. Wabnitz, H. Weise, R. Wichmann, K. Witte, A. Wolf, M. Wulff, and M. Yurkov, ``The European X-Ray Free-Electron Laser Technical design report," (DESY, 2007).



\bibitem{palffy08} A. P\'{a}lffy, J. Evers, and C. H. Keitel, ``Electric-dipole-forbidden nuclear transitions driven by super-intense laser fields,'' Phys. Rev. C \textbf{77}, 044602 (2008).



\bibitem{baranov14}D. G. Baranov, A. P. Vinogradov, and A. A. Lisyansky,
``Abrupt Rabi oscillations in a superoscillating electric field,'' Opt. Lett. \textbf{39}, 6316--6319 (2014).
\bibitem{ber15}R. Ber and M. Schwartz, ``Unusual transitions made possible by superoscillations,'' arXiv:1502.01406 (2015).
\bibitem{kempf15}A. Kempf and A. Prain, ``Driving quantum systems with superoscillations,''
arXiv:1510.04353 (2015).


\bibitem{eastham73}M. S. P. Eastham, \emph{The Spectral Theory of Periodic
Differential Equations} (Scottish Academic, 1973).



\bibitem{breuer02} {H. P. Breuer and F. Petruccione, \textit{The Theory of Open Quantum System} (Oxford University, 2002).}


\bibitem{rabi36} I. I. Rabi, ``On the process of space quantization,'' Phys. Rev. \textbf{49}, 324 (1936).
\bibitem{rabi37} I. I. Rabi, ``Space quantization in a gyrating magnetic field,'' Phys. Rev. \textbf{51}, 652 (1937).
\bibitem{solano11}E. Solano, ``Viewpoint: The dialogue between quantum light and matter,'' Physics \textbf{4}, 68 (2011).
\bibitem{albert12}V. V. Albert, ``Quantum Rabi model for N-state atoms,''  Phys. Rev. Lett. \textbf{108}, 180401 (2012).
\bibitem{hwang15}M. J. Hwang, R. Puebla, and M. B. Plenio, ``Quantum phase transition and universal dynamics in the Rabi model,''  Phys. Rev. Lett. \textbf{115}, 180404 (2015).


\bibitem{braak11}D. Braak, ``Integrability of the Rabi model,'' Phys. Rev. Lett. \textbf{107}, 100401 (2011).
\bibitem{chen12} {Q. H. Chen, C. Wang, S. He, T.Liu, and K. L. Wang, ``Exact solvability of the quantum Rabi model using Bogoliubov operators,'' Phys. Rev. A \textbf{86}, 023822 (2012).}

\bibitem{jaynes63} E. T. Jaynes and F. W. Cummings, ``Comparison of quantum and semiclassical radiation theories with application to the beam maser,'' Proc. IEEE \textbf{51}, 89--109 (1963).


\bibitem{shevchenko10} S. N. Shevchenko, S. Ashhab, and F. Nori, ``Landau-Zener-St¨¹ckelberg interferometry,'' Phys. Rep. \textbf{492}, 1--30 (2010).
\bibitem{huang11} P. Huang, J. Zhou, F. Fang, X. Kong, X. Xu, C. Ju, and J. Du, ``Landau-Zener-St¨¹ckelberg interferometry of a single electronic spin in a noisy environment,'' Phys. Rev. X \textbf{1}, 011003 (2011).
\bibitem{ferron12} A. Ferr\'{o}n, Daniel Dom\'{\i}nguez and M. S\'{a}nchez, ``Tailoring population inversion in Landau-Zener-St¨¹ckelberg interferometry of flux qubits,'' Phys. Rev. Lett. \textbf{109}, 237005 (2012).
\bibitem{cao13} G. Cao, H. O. Li, T. Tu, L. Wang, C. Zhou, M. Xiao, G. C. Guo, H. W. Jiang, and G. P. Guo, ``Ultrafast universal quantum control of a quantum-dot charge qubit using Landau-Zener-St¨¹ckelberg interference,'' Nat. Commun. 4, 1401 (2013).
\bibitem{dupont13} E. Dupont-Ferrier, B. Roche, B. Voisin, X. Jehl, R. Wacquez, M. Vinet, M. Sanquer, and S. De Franceschi, ``Coherent coupling of two dopants in a Silicon nanowire probed by Landau-Zener-St¨¹ckelberg interferometry,'' Phys. Rev. Lett. \textbf{110}, 136802 (2013).


\bibitem{izmalkov04} A. Izmalkov, M. Grajcar, E. Il'ichev, N. Oukhanski, T. Wagner, H. G. Meyer, W. Krech, M. H. S. Amin, A. Maassen van den Brink, and A. M. Zagoskin, ``Observation of macroscopic Landau-Zener transitions in a superconducting device,'' Europhys. Lett. \textbf{65}, 844 (2004).
\bibitem{oliver05} W. D. Oliver, Y. Yu, J. C. Lee, K. K. Berggren, L. S. Levitov, and T. P. Orlando, ``Mach-Zehnder interferometry in a strongly driven superconducting qubit,'' Science \textbf{310}, 1653--1657 (2005).
\bibitem{sillanpaa06} M. Sillanp\"a\"a, T. Lehtinen, A. Paila, Y. Makhlin, and P. Hakonen, ``Continuous-time monitoring of Landau-Zener interference in a Cooper-pair box,'' Phys. Rev. Lett. \textbf{96}, 187002 (2006).
\bibitem{izmalkov08} A. Izmalkov, S. H. W. van der Ploeg, S. N. Shevchenko, M. Grajcar, E. Il'ichev, U. H\"ubner, A. N. Omelyanchouk, and H. G. Meyer, ``Consistency of ground state and spectroscopic measurements on flux qubits,'' Phys. Rev. Lett. \textbf{101}, 017003 (2008).
\bibitem{lahaye09} M. D. LaHaye, J. Suh, P. M. Echternach, K. C. Schwab, and M. L. Roukes, ``Nanomechanical measurements of a superconducting qubit,'' Nature \textbf{459}, 960--964 (2009).
\bibitem{johansson09} J. Johansson, M. H. S. Amin, A. J. Berkley, P. Bunyk, V. Choi, R. Harris, M. W. Johnson, T. M. Lanting, Seth Lloyd, and G. Rose, ``Landau-Zener transitions in a superconducting flux qubit,'' Phys. Rev. B \textbf{80}, 012507 (2009).
\bibitem{silveri15} M. P. Silveri, K. S. Kumar, J. Tuorila, J. Li, A. Veps\"{a}l\"{a}inen, E. V. Thuneberg, and G. S. Paraoanu, ``St\"uckelberg interference in a superconducting qubit under periodic latching modulation,'' New J. Phys. \textbf{17}, 043058 (2015).



\bibitem{thaichroen15} N. Thaicharoen, A. Schwarzkopf, and G. Raithel, ``Measurement of the van der Waals interaction by atom trajectory imaging,'' Phys. Rev. A \textbf{92}, 040701 (2015).

\bibitem{liao11} W. T. Liao, A. P\'{a}lffy, and C. H. Keitel, ``Nuclear coherent population transfer with X-ray laser pulses,'' Phys. Lett. B \textbf{705}, 134--138 (2011).

\bibitem{crespi12}A. Crespi, S. Longhi, and R. Osellame, ``Photonic realization of the quantum Rabi model,'' Phys. Rev. Lett. \textbf{108}, 163601 (2012).

\bibitem{adams13} B. W. Adams, C. Butha, S. M. Cavalettob, J. Eversb, Z. Harmanbc, C. H. Keitelb, A. P\'{a}lffyb, A. Pic\'{o}na, R. R\"ohlsbergerd, Y. Rostovtseve, and K. Tamasakuf, ``X-ray quantum optics,'' J. Mod. Opt. \textbf{60}, 2--21 (2013).



\end{thebibliography}
\end{document}